\documentclass[useAMS,usenatbib]{mn2e}
\usepackage{amsmath}
\usepackage{graphicx}
\usepackage{color}
\usepackage{ulem}
\usepackage{epsf}
\def\msun{{\rm M_{\odot}}}

\def\me{{\dot M_{\rm Edd}}}
\def\le{{L_{\rm Edd}}}

\title[Pulsing ULXs: tip of the iceberg?] {Pulsing ULXs: tip of the iceberg?}

\author[Andrew King,  Jean--Pierre Lasota and  W\l{}odek Klu\'zniak]
{Andrew King$^{1, 2, 3, 4}$,  Jean--Pierre Lasota$^{4, 5}$ and W\l{}odek Klu\'zniak$^{5}$\\
$^{1}$ Theoretical Astrophysics Group, Department of Physics \& Astronomy, University of Leicester, Leicester LE1 7RH, UK\\
$^{2}$ Astronomical Institute Anton Pannekoek, University of Amsterdam, Science Park 904, 1098 XH Amsterdam, Netherlands\\
$^{3}$ Leiden Observatory, Leiden University, Niels Bohrweg 2, NL-2333 CA Leiden, Netherlands\\
$^{4}$ Institut d'Astrophysique de Paris, CNRS et Sorbonne Universit\'es, UPMC Paris~06, UMR 7095, 98bis Bd Arago, 75014 Paris, France\\  
$^{5}$ Nicolaus Copernicus Astronomical Center, Polish Academy of Sciences, ul. Bartycka 18, 00-716 Warsaw, Poland\\     
}

\date{\today}

\volume{000}

\pagerange{\pageref{firstpage}--\pageref{lastpage}} \pubyear{2017}

\begin{document}

\label{firstpage}

\maketitle

\begin{abstract}
We consider the three currently known pulsing ultraluminous X--ray sources (PULXs). We show that in one of them the observed spinup rate requires super--Eddington accretion rates at the magnetospheric radius, even if magnetar--strength fields are assumed. In the two other systems a normal--strength neutron star field implies super--Eddington accretion at the magnetosphere. Adopting super--Eddington mass transfer as the defining characteristic of ULX systems, we find the parameters required for self--consistent simultaneous fits of the luminosities and spinup rates of the three pulsed systems. These imply near--equality between their magnetospheric radii $R_M$ and the spherization radii $R_{\rm sph}$ where radiation pressure becomes important and drives mass loss from the accretion disc. We interpret this near--equality as a necessary condition for the systems to appear as pulsed, since if it is violated the pulse fraction is small. We show that as a consequence all PULXs must have spinup rates
$\dot\nu \ga 10^{-10}\, {\rm s^{-2}}$, an order of magnitude higher than in any other pulsing neutron--star binaries. The fairly tight conditions required for ULXs to show pulsing support our earlier suggestion that many unpulsed ULX systems must actually contain neutron stars rather than black holes.
\end{abstract}

\begin{keywords}
  accretion, accretion discs -- binaries: close -- X-rays: binaries --
  black hole physics -- neutron stars -- pulsars: general  
 \end{keywords}

\section{Introduction}

Ultraluminous X--ray sources (ULXs) have apparent luminosities $L$ higher than the Eddington limit for a standard stellar--mass accretor (typically $L \ga 10^{39}\,{\rm erg\, s^{-1}}$), but are clearly not supermassive. One suggestion is that they contain intermediate--mass black holes \citep[IMBH:][]{CM99}, but by now there is evidence that most -- or perhaps all -- ULXs are standard X--ray binaries in some unusual evolutionary phase, probably characterized by super--Eddington mass transfer rates \citep{Kingetal01}. 

It is often assumed that the accretor in ULXs is always a black hole, 
although \citet{Kingetal01} pointed that neutron--star and white--dwarf ULXs are possible. 
There are now three ULX systems (M82 X--2, \citet{Bachettietal14},   NGC 7793 P13, \citet{Furstetal16,Israeletal17}, and NGC5907 ULX1, \citet{Israeletal16}) which show regular pulses with periods $\sim 1$\,s, characteristic of neutron--star accretors (see Table 1 for a list of observed properties). One type of explanation of these systems \citep[cf][]{Tong15,Eksi15,Dosso15} is that they are magnetars, i.e. with very high magnetic fields, so that the reduction of the electron scattering cross section below the Thomson value allows super--Eddington luminosities in certain directions. Here we first show that the observed spinup rates for ULX pulsars (hereafter denoted as PULXs) imply strongly super--Eddington accretion in one of the three known systems, even if magnetar--strength fields are assumed.
 
This result supports the conclusions of \citet[][hereafter KL16]{KL16}, who showed instead that M82 X--2 fits naturally into a unified picture applying to all ULXs as X--ray binaries with beamed emission caused by super--Eddington mass transfer rates \citep{Kingetal01}. In the rest of this paper we consider whether this is true of the other two recently--discovered pulsing ULXs. Our results show that this is possible.
They also suggest that the condition that ULXs should show detectable pulses is quite restrictive, in particular requiring very high spin--period derivatives, as indeed observed. This in turn reinforces the
conclusion of  KL16 that many non--pulsing ULXs  usually assumed to contain black holes in fact have neutron--star accretors.

\begin{center}
\begin{table*}
\centerline{\bf Table 1: observed properties of PULXs}
\bigskip
{\small
\hfill{}
\begin{tabular}{ |l|c|c|c| } 
 \hline\hline
 Name & M82 ULX2$^1$ & NGC 7793 P13$^2$  & NGC5907 ULX1$^3$\\ 
 \hline\hline
 $L_X (\rm max)$  [erg\,s$^{-1}$] & $1.8 \times 10^{40}$  & $5\times 10^{39}$ & $\sim 10^{41}$\\ 
 \hline
 $P_s $ [s] & 1.37 &  0.42& 1.13 \\ 
 \hline
 $\dot \nu$ [s$^{-2}$]& $ 10^{-10}$ & $4 \times 10^{-11}$ & $4 \times 10^{-9}$\\ 
 \hline
 $P_{\rm orb}$ [d]& 2.51 (?) & 64 & 5.3(?) \\ 
 \hline
 $M_2$ [$\rm M_{\odot}$]& $\ga 5.2$ &18--23  &\\ 
 \hline\hline
\end{tabular}}
\hfill{}
\label{tb:ulx1}\\
\vskip 0.2truecm 
$^1$\citet{Bachettietal14}, $^2$\citet{Furstetal16,Israeletal17,Motchetal14}; Pietrzy\'nski (2016, private communication),\\ $^3$\citet{Israeletal16} 
\end{table*}
\end{center}
\begin{center}%[ht]
\begin{table*}%[ht]
\centerline{\bf Table 2: minimum Eddington accretion factors for PULXs required by the observed spinup rates}
\bigskip
{\small
\hfill{}
\begin{tabular}{ |l|c|c|c| } 
 \hline\hline
 Name & M82 ULX2 & NGC 7793 P13  & NGC5907 ULX1\\ 
 \hline\hline
 $\dot m(R_M)q^{7/12}$; $\mu_{30}=1$  & 5.8  & 2.0 & 429\\ 
 \hline
 $\dot m(R_M)q^{7/12}$; $\mu_{30}=1000$ & 0.6 &  0.2 & 43 \\ 
\hline\hline
\end{tabular}}
\hfill{}
\label{tb:ulx2}\\
\vskip 0.2truecm 
\end{table*}
\end{center}
\section{Spin and Spinup}
The physics of accretion on to a magnetic neutron star is determined by its Alfv\'en radius $R_M$, where the matter stresses in the accretion disc roughly balance the magnetic stresses specified by the dipole moment $\mu = 10^{30}\mu_{30}\,{\rm G\,cm^{3}}$. This gives
\begin{equation}
R_M = 2.6 \times 10^8 q\dot M_{17}^{-2/7}m_1^{-1/7}\mu_{30}^{4/7}\,{\rm cm},
\label{rm}
\end{equation}
Here $q \la 1$ is a factor taking into account the geometry of the accretion flow at the magnetosphere (often taken $= 0.5$ for geometrically thin discs) and $\dot M_{17}$ is the accretion rate at $R_M$ in units of $10^{17}\,{\rm g\,s^{-1}}$ \citep[cf e.g.][hereafter FKR02]{FKR02}; $m_1$ is the accretor mass in $\msun$. Within $R_M$, disc material is assumed to flow along fieldlines. 

The disc angular momentum arriving at $R_M$ predicts a theoretical maximum spinup rate 
\begin{equation}
\dot\nu = 3.1\times 10^{-12}q^{1/2}\dot M_{17}^{6/7}m_1^{3/7}\mu_{30}^{2/7}I_{45}^{-1}{\rm \,s^{-2}}
\label{nudot}
\end{equation}
valid for spin frequencies $\nu$ slower than the equilibrium value, i.e. spin periods longer than specified by the quantity $P_{\rm eq}$ defined in (\ref{peq}) below. 
Here  $I_{45}$ is the neutron star moment of inertia in units of $10^{45}{\rm \,g\,cm^2}$ (see e.g. FKR02; note that in KL16 this equation was given with a spurious extra factor $R_6^{6/7}(m_1)^{-6/7}$ in KL16: this had only a minimal effect on the results of that paper as it adopted $R_6 = 1, m_1 = 1.4$; $R_6$ is the neutron star radius in units of $10^6$\,cm).

Using the observed values of $\dot\nu$ (Table 1) in equation (\ref{nudot}) and taking $m_1=1.4$, $I_{45}=1$ gives the minimum accretion rate at $R = R_M$ in Eddington units as 
\begin{equation}
\dot m(R_M) = \frac{\dot M(R_M)}{\me} = 5.8\left(\frac{\dot\nu_{-10}}{q^{1/2}}\right)^{7/6}\mu_{30}^{-1/3}
\label{super}
\end{equation}
where $\dot\nu_{-10} = \dot\nu/10^{-10}$ and we have taken the Eddington rate for an accreting neutron star as $\me =1.6\times 10^{18}\,{\rm g\,s^{-1}}$. Table 2 shows that all three systems must have 
super--Eddington mass transfer rates if $\mu_{30} = 1$, as typical for neutron stars, and that two of the
three systems would still be super--Eddington even for magnetar--strength fields $\mu_{30} = 10^3$.
This suggests that super--Eddington accretion is the defining characteristic of all ULXs, pulsed or unpulsed, 
removing the need to invoke special magnetic mechanisms to explain the former group.

\section{ULX Accretion}
\begin{table*}
\centerline{\bf Table 3: derived properties of PULXs}
\bigskip
{\small
\hfill{}
\begin{tabular}{ |c|c|c|c| } 
 \hline\hline
 Name & M82 ULX2 & NGC 7793 P13  & NGC5907 ULX1\\ 
 \hline\hline
  $\dot m_0$ & 36  & 20 & 91\\ 
 \hline
 $\mu\, q^{7/4}m_1^{-1/2}I_{45}^{-3/2}$ [Gcm$^3$]& $9.0\times 10^{28}$ & $2.3\times 10^{28}$ & $2.3\times 10^{31}$ \\ 
  \hline
 $R_{\rm sph}m_1^{-1}$ [cm]& $5.9\times 10^7$ & $3.3\times 10^7$ & $1.3\times 10^8$ \\
 \hline
$R_M m_1^{-1/3}I_{45}^{-2/3}$ [cm] &$1.6\times 10^7$ & $8.7\times 10^6$&$1.9\times 10^8 $\\
\hline
$R_{\rm co}m_1^{-1/3}$[cm]& $1.9\times 10^8$ &  $8.4 \times 10^8$ & $1.6 \times 10^8$\\
\hline
$P_{\rm eq}q^{-7/6}m_1^{1/3}$ [s]& 0.09 & 0.02 & 1.86\\ 
\hline
 $t_{\rm eq}$ [yr]& 1647 & 40776  & 0 \\ 
\hline\hline
\end{tabular}}
\hfill{}
%\caption{Model parameters of ULX pulsars}
\label{tb:ulx3}
\end{table*}

Motivated by the conclusion of the last Section,
our aim is to see how pulsed systems fit into the general picture of ULXs as binaries with super--Eddington mass supply rates and consequent collimated (or beamed) emission. We adopt the equations used by KL16 to study M82 ULX2, the first PULX discovered \citep{Bachettietal14}. 

\citet{SS73} studied what happens if mass is transferred to a compact accretor at a 
super--Eddington rate $\dot M_0 =\dot m_0\me$, where $\dot m_0 >1$, and $\me$ is the rate which 
would produce
the Eddington luminosity if it reached the accretor. They reasoned that the disc would remain stable outside the spherization radius
\begin{equation}
R_{\rm sph} \simeq \frac{27}{4}\dot m_0 R_g \simeq 1\times 10^6\dot m_0m_1\, {\rm cm},
\label{rsph}
\end{equation}
where $R_g = GM/c^2$ is the gravitational radius of the accretor. $R_{\rm sph}$ is close to the trapping radius \citep[see e.g.,][]{Begelman06,Poutanenetal07}.
At this point the accretion luminosity is close to the local Eddington value, so we can expect significant outflow here, and from all disc radii $\la R_{\rm sph}$ also. To prevent the emission at each
disc radius within $R_{\rm sph}$ exceeding
its local Eddington limit, the outflow must arrange that the accretion rate through the disc decreases as
\begin{equation}
\dot M(R) \simeq \frac{R}{R_{\rm sph}}\dot m_0\me.
\label{mr}
\end{equation}
Then the total accretion luminosity is \citep{SS73}
\begin{equation}
L \simeq \le\left[1 + \ln\dot m_0\right].
\label{ldisc}
\end{equation}

The outflow from the disc is likely to be quasispherical and scatter the emission from the disc, but must have narrow evacuated funnels along the disc axis where radiation can escape freely. Using a combination of observational and theoretical arguments, \citet{King09} gives an approximate formula
\begin{equation}
b \simeq \frac{73}{{\dot m_0}^2}
\label{beam}
\end{equation}
for the total beam solid angle $4\pi b$, valid for $\dot m_0 \ga \sqrt{73} \simeq 8.5$. This form reproduces the inverse luminosity--temperature correlation $L_{\rm soft}\sim T^{-4}$
observed for soft X--ray components in ULX spectra \citep[cf][]{KajavaP09}, and 
\citet{Mainierietal10} find that it gives a good representation of the local luminosity function of ULXs. 

The apparent (isotropic) luminosity $L_{\rm sph} = L/b$ for a given $\dot m_0$ now follows from 
(\ref{ldisc}, \ref{beam}) as
\begin{equation}
\frac{m_1}{L_{40}} \simeq \frac{4500}{{\dot m_0}^2(1 + \ln\dot m_0)}
\label{mL}
\end{equation}
\citep{King09}, where $L_{40}$ is the apparent luminosity in units of $10^{40}\, {\rm erg\, s^{-1}}$. We can use this and the observed $L_{40}$ to find $\dot m_0$ for the three systems of Table 1 
%%%%%%%%%%%%
\citep[data from][]{Bachettietal14,Furstetal16,Israeletal17,Motchetal14,Israeletal16}). 
%%%%%%%%%%%%
This in turn gives the
the mass transfer rates $\dot M_0$ in these binaries (Table 3), which are of order  
$\dot M_0 \sim 0.54 - 2\times 10^{-6}\msun\,{\rm yr}^{-1}$.
Self--consistency requires that (\ref{mr}) must hold, so that
\begin{equation}
\frac{R_{\rm sph}}{R_M} = \frac{\dot M_0}{\dot M(R_M)},
\label{consist}
\end{equation}
which gives the value of $\mu$ for each system (Table 3). This now allows us to define the equilibrium spin period $P_{\rm eq}$ as the Kepler period at the magnetospheric radius (e.g. FKR02), i.e.
\begin{equation}
P_{\rm eq} = 2\pi\left(\frac{R_M^3}{GM}\right)^{1/2} \simeq 3 q^{3/2}\dot M_{17}^{-3/7}m_1^{-5/7}\mu_{30}^{6/7}\, {\rm s}
\label{peq}
\end{equation}
It is usually assumed that the disc gas cannot overcome the centrifugal barrier to spin up the neutron star to shorter periods than this. We can find the time $t_{\rm eq}$ for each pulsar to reach its equilibrium spin at its current spinup rate from
\begin{equation}
t_{\rm eq} = \frac{1}{\dot\nu}\left(\frac{1}{P_{\rm eq}} -\frac{1}{P}\right).
\label{teq}
\end{equation}
Combining equations (\ref{rm}, \ref{nudot}, \ref{rsph}, \ref{consist}) gives
\begin{eqnarray}
&& \dot M_{17}(R_M) = 390\, q^{7/9}m_1^{-8/9}\mu^{4/9}_{30}  \label{mdotrm}\\
 && R_M = 1.6\times 10^7 \dot\nu_{-10}^{2/3}m_1^{1/3}I_{45}^{2/3}\,{\rm cm} \label{RM}\\
 && \mu_{30} = 0.09\,q^{-7/4}\dot\nu_{-10}^{3/2}m_1^{1/2}I_{45}^{3/2}
\label{mu30}
\end{eqnarray}
where $\dot\nu_{-10} = \dot\nu/10^{-10}\,{\rm s}^{-2}$.

Table 3 gives the derived values of $R_M$ and 
$R_{\rm sph}$ for the three currently--known systems. Both radii are within the corotation radius $R_{\rm co}\equiv\left(GMP_s^2/4\pi^2\right)^{1/3}$ for each system.
From equations (\ref{peq}) and (\ref{mdotrm}) one gets the equilibrium spin period
\begin{equation}
P_{\rm eq}=0.23 q^{7/6}m_1^{-1/3}\mu_{30}^{2/3}.
\label{peq2}
\end{equation}

\section{Pulsing ULXs}
Remarkably, Table 3 show that all three PULXs have $R_{\rm sph}\simeq R_M$, despite these systems having rather different parameters. 
For NGC 5907 ULX1, a slight increase of $m_1$ above 1 or a decrease of $q$ below 1 is needed to make $R_{\rm sph}$ formally larger than
$R_M$, as we assumed above in taking $\dot M(R) \propto R$. The small difference between $R_{\rm sph}$ and $R_M$ means that the flow is strongly super--Eddington on reaching $R_M$. Most of this cannot land on the neutron star (let alone its polecaps only) and so must be ejected.
Since accretion within the magnetosphere is highly dissipative (flow along fieldlines requires a hypersonic flow to bend and shock) the reasoning of \citet{SS73} suggests a similar scaling for this flow also. The resulting outflow should lead to a qualitatively similar beaming effect, accounting for the ULX behaviour, although we cannot now rely on parameters derived assuming the original beaming formula (\ref{beam}) as most of the flow is now magnetospheric. 
We note that our model does not have to assume a special geometry for the pulses to be seen.The disc-based funnel is wide, when compared with the size of the neutron star.
To see the ultra-luminous radiation at all, and the pulses, one has to be looking down the funnel, of course, but otherwise one expects pulses generically. One does not have to be observing the pulsed emission from the neutron star directly. The pulsar beam sweeps along the interior surface of the funnel, and it is enough to see a part of that surface as one peers down the opening \citep[see also][]{BS76}.
But we should ask why it appears that the three known pulsing ULXs all have $R_{\rm sph}\simeq R_M$, since it seems physically perfectly possible to arrange instead that $R_{\rm sph} \gg R_M$.

It appears very likely that this is a selection effect: if $R_{\rm sph} \gg R_M$, the pulsed fraction of the emitted luminosity would probably be very small. The emission from the accreting magnetic polecaps is $\la p\le$, where $p$ is the fractional polecap area,  while the unpulsed emission is 
$\sim \le\left[1 + \ln\dot m_0\right] \gg p\le$. Both components would probably be beamed by the disc outflow (i.e. by the factor $b$, cf eqn \ref{beam}), so to make the pulsed emission noticeable would require very tight beaming within the magnetosphere.  This reasoning suggests that instead pulsing is only detectable provided that $R_M$ is not much smaller than $R_{\rm sph}$, i.e. 

\bigskip
\noindent
{\sl a pulsing ULX system must have} 
\begin{equation}
R_{M} \sim  f R_{\rm sph},\,f\sim 0.3 - 1
\label{puls1}
\end{equation}
and so from (\ref{mr}) 
\begin{equation}
\dot M_0 \ga \dot M(R_M) \ga f\dot M_0.
\label{puls2}
\end{equation}

\noindent
We see from (\ref{rsph}, \ref{RM}) that this requires
\begin{eqnarray}
&&f\dot M(R_M) \sim 3.9\times 10^{19}q^{7/9}m_1^{-8/9}\mu_{30}^{4/9}\,{\rm g\,s^{-1}} \\
&& \ \ \ \ \ \ \ \ \ \ \ \ \ \sim 24 q^{7/9}\me m_1^{-17/9}\mu_{30}^{4/9}
\label{puls3}
\end{eqnarray}
and from this equation and (\ref{nudot}) that
\begin{equation}
\dot\nu = 5.2\times 10^{-10}q^{5/6}m_1^{-1/3}\mu_{30}^{6/7}I_{45}^{-1}\,{\rm s^{-2}}.
\label{puls4}
\end{equation}
This result explains why the the spin derivatives of the PULXs are all more than a factor 10 larger than for any other pulsing neutron star systems \citep[cf][]{KL15}, provided $\mu_{30} \ga 0.1$. We see that from (\ref{puls3}) that 
mass transfer rates $\dot M_0 \sim 50\me \sim 5\times 10^{-7}\msun\,{\rm yr^{-1}}$ are needed. 
Given the very short spinup timescales $t_{\rm eq}$ it seems very unlikely that we observe these systems during their only approach to spin equilibrium. Instead they are all probably close to $P_{\rm eq}$
with alternating spinup and spindown phases. Evidently we can only see these systems during spinup phases (so that $\dot\nu$ has its maximum value) because centrifugal repulsion during spindown presumably reduces the accretion rate and so the luminosity. 
\section{Conclusions}

We have seen that a ULX system containing a magnetic neutron star can apparently show pulses only under rather special conditions which make the magnetospheric radius $R_M$ of similar size to the spherization radius $R_{\rm sph}$. We note that this is highly likely to give the rather sinusoidal pulse profiles observed 
for these systems as well as their high observed spinup rates.

In addition to the requirement (\ref{puls1}) a further effect makes such systems inherently rare.
A  simple way in which pulsing can fail is that adding even a small amount of mass to a neutron star
is apparently able to weaken the surface field significantly. In this case matter accretes axisymmetrically on to the neutron star and there is no pulsing.

The combination of these effects with the requirements found in the previous Section 
mean that ULXs containing magnetic neutron stars are likely to spend only a relatively  short fraction of their ULX phase emitting observable pulses.  We conclude (as in KL16) that it is likely that a significant fraction of ULXs actually have neutron--star accretors rather than the black holes usually assumed. As we remarked in that paper, this is not surprising in view of the fact that for a given binary mass transfer rate, neutron--star systems are more super--Eddington (and so if eqn \ref{beam} holds, more beamed) than black--hole systems.

 \section*{Acknowledgments}

JPL acknowledges support from the French Space Agency CNES. This research was supported by the Polish NCN grants No. 2013/08/A/ST9/00795, 2012/04/A/ST9/00083 and 2015/19/B/ST9/01099. ARK and WK thank the Institut d'Astrophysique, Paris, for a visit during which this work was performed. Theoretical astrophysics research at the University of Leicester is supported by an STFC Consolidated Grant.

\bsp

\label{lastpage}

\end{document}